\documentclass[10pt, onecolumn]{IEEEtran}
\IEEEoverridecommandlockouts
\usepackage[nopostdot,nogroupskip,style=super,nonumberlist,acronym]{glossaries}
\usepackage{graphicx}
\usepackage[utf8]{inputenc}
\usepackage[dvipsnames]{xcolor}
\usepackage[doublespacing]{setspace}
\usepackage{booktabs}
\usepackage{subfig}
\usepackage{hyperref}
\usepackage{amssymb}
\usepackage{algorithm}
\usepackage{algorithmic}
\usepackage{chngcntr}
\counterwithin{figure}{section}
\newacronym{3gpp}{3GPP}{3rd Generation Partnership Project}
\newacronym{ai}{AI}{artificial intelligence}
\newacronym{aoi}{AoI}{age of information}
\newacronym{aiaas}{AIaaS}{\gls{ai} as-a-service}
\newacronym{ap}{AP}{access point}
\newacronym{awgn}{AWGN}{additive white Gaussian noise}
\newacronym{bs}{BS}{base station}
\newacronym{cdf}{CDF}{Cumulative Distribution Function}
\newacronym{e2e}{E2E}{end-to-end}
\newacronym{edrs}{EDRS}{European Data Relay System}
\newacronym{fcc}{FCC}{Federal Communications Commission}
\newacronym{geo}{GEO}{geostationary orbit}
\newacronym{fso}{FSO}{free-space optical}
\newacronym{gs}{GS}{ground station}
\newacronym{gsl}{GSL}{ground-to-satellite link}
\newacronym{hap}{HAP}{high-altitude platform}
\newacronym{iot}{IoT}{Internet of Things}
\newacronym{irs}{IRS}{intelligent reflecting surfaces}
\newacronym{isl}{ISL}{inter-satellite link}
\newacronym{kpi}{KPI}{key performance indicator}
\newacronym{leo}{LEO}{low Earth orbit}
\newacronym{los}{LoS}{line-of-sight}
\newacronym{lpwan}{LPWAN}{low-power wide-area network}
\newacronym{mcs}{MCS}{modulation and coding scheme}
\newacronym{meo}{MEO}{medium Earth orbit}
\newacronym{mimo}{MIMO}{multiple-input multiple-output}
\newacronym{nbiot}{NB-IoT}{Narrowband \gls{iot}}
\newacronym{ngso}{NGSO}{non-geostationary orbit}
\newacronym{nr}{NR}{New Radio}
\newacronym{ofdm}{OFDM}{orthogonal frequency division multiplexing}
\newacronym{paa}{PAA}{point-ahead-angle}
\newacronym{qos}{QoS}{quality of service}
\newacronym{rf}{RF}{radio frequency}
\newacronym{rtt}{RTT}{round-trip time}
\newacronym{ue}{UE}{user equipment}
\newacronym{snr}{SNR}{signal-to-noise ratio}
\newacronym{uav}{UAV}{unmanned aerial vehicle}
\newacronym{dtn}{DTN}{Delay and Disruption Tolerant Networking}
\makeglossaries

\newcommand{\rev}[1]{\textcolor{black}{#1}}

\author{\IEEEauthorblockN{Israel Leyva-Mayorga\IEEEauthorrefmark{1}, Beatriz Soret\IEEEauthorrefmark{1}\IEEEauthorrefmark{2},  Bho Matthiesen\IEEEauthorrefmark{3}\IEEEauthorrefmark{4}, Maik R{\"o}per\IEEEauthorrefmark{3},\\Dirk W{\"u}bben\IEEEauthorrefmark{3}, Armin Dekorsy\IEEEauthorrefmark{3}, and Petar Popovski\IEEEauthorrefmark{1}\IEEEauthorrefmark{4}}

\IEEEauthorblockA{\IEEEauthorrefmark{1}Department of Electronic Systems, Aalborg University, Aalborg, Denmark}

\IEEEauthorblockA{\IEEEauthorrefmark{1}Telecommunications Research Institute (TELMA), University of Malaga, Malaga, Spain}

\IEEEauthorblockA{\IEEEauthorrefmark{3}Gauss-Olbers Center, c/o University of Bremen, Dept. of Communications Engineering, Germany}

\IEEEauthorblockA{\IEEEauthorrefmark{4}University of Bremen, U Bremen Excellence Chair, Dept.\ of Communications Engineering, Germany}

\IEEEauthorblockA{Email: \{ilm, bsa, petarp\}@es.aau.dk, \{matthiesen, roeper,wuebben, dekorsy\}@ant.uni-bremen.de}}
\begin{document}
%\rhbooktitle{NGSO Communications Systems}
%\markboth{NGSO Communications Systems}{NGSO Constellation Design for Global Connectivity}

%\author{Israel Leyva-Mayorga^1, Beatriz Soret^1,  Bho Matthiesen^{2,3}, Maik R{\"o}per^2, Dirk W{\"u}bben^2, Armin Dekorsy^2, and~Petar~Popovski^1^3\thanks{^1 Department of Electronic Systems, Aalborg University, Aalborg, Denmark}\thanks{Gauss-Olbers Center, c/o University of Bremen, Dept. of Communications Engineering, Germany} \thanks{University of Bremen, U Bremen Excellence Chair, Dept.\ of Communications Engineering, Germany}}

\title{NGSO Constellation Design for Global Connectivity}

\maketitle
%\ilmn{Extent: Approximately 6,600--8,800 words / 15--20 pages, based on 550 words per full-text page with a 20\% allowance for figures}

%\ilmn{Deadline: October 1 (flexible)}

\setlength{\glsdescwidth}{0.8\textwidth}
\clearpage\begingroup%\let\newpage\relax
\renewcommand{\arraystretch}{0.9}
\printglossary[type=\acronymtype]\endgroup

%\rhbooktitle{NGSO Communications Systems}
%\markboth{NGSO Communications Systems}{NGSO Constellation Design for Global Connectivity}
\section{Introduction}
\label{sec:intro}
Providing global connectivity is not possible with terrestrial infrastructure alone. \rev{This is due to a multitude of factors, the most important of which are geographical conditions and economic reasons. So far, it seems that every new mobile wireless generation has ambitions to connect sparsely populated areas, but terrestrial options have not proven to be cost-effective.} While providing radio access merely necessitates the deployment of a \gls{bs} or \gls{ap} in the area of interest, connecting this infrastructure to the core network and, hence, to the Internet through \emph{backhaul} and, possibly, \emph{fronthaul} links is much more challenging. A clear use case is providing global connectivity to vessels in open ocean, where deploying \glspl{bs} and the necessary backhaul links (i.e., sea cables) to the many possible routes is not feasible. 

%Many wired and wireless communication technologies have appeared and matured in recent years, increasing the performance, resource efficiency, and communication range. Nevertheless, the only one that has proven to be sufficient to achieve global connectivity is satellite communications. \bs{Are you thinking of UAVs? I'd remove this paragraph, satcoms are mature enough and need to further justification here}. 

In contrast, \gls{geo} satellites have been used to provide global communication coverage for several decades, for instance, for TV broadcasting or maritime connectivity. In addition, GPS is an example of a widespread \gls{meo} service. Even though \gls{geo} satellites are able to provide global service availability in underserved and disconnected areas~\cite{3GPPTR38913},  they are not efficient on their own as a competitive global connectivity solution. Due to the high altitude of the orbit, \gls{geo} satellites suffer from a long propagation delay and a high signal attenuation. The latter aspect is specially problematic when devices with energy and size restrictions attempt to communicate in the uplink, which are some of the defining characteristics of \gls{iot} devices~\cite{Qu2017,Liberg2020}. 

\Gls{ngso} satellite constellations represent a cornerstone in the NewSpace paradigm and thus have become one of the hottest topics for the industry, academia, but also for national space agencies and regulators. \rev{For instance, numerous companies worldwide, including Starlink, OneWeb, Kepler, SPUTNIX, and Amazon have started or will soon start to deploy their own \gls{ngso} constellations~\cite{Di2019}, which aim to provide either broadband~\cite{Su2019, DelPortillo2019} or \gls{iot} services~\cite{Qu2017}.} One of the major drivers for such a high interest on \gls{ngso} constellations is that, with an appropriate design, they are capable of providing global coverage and connectivity. %\ilm{Furthermore, \gls{ngso} constellations are greatly flexible and can be designed to be deployed alone or in combination with satellites deployed at higher orbits~\cite{Chien2019}}. 
 While global connectivity can also be provided by a small set of \gls{geo} satellites, \Gls{ngso} constellations present three main advantages over terrestrial and \gls{geo} satellite communications: 
\begin{enumerate}
    \item \textbf{Short propagation delay:} Electromagnetic waves propagate faster in the vacuum than in optic fiber, which has typical refraction index of $1.44$ to $1.5$~\cite{Handley2018}. Moreover, \gls{ngso} satellites are deployed at much lower altitudes than \gls{geo} satellites, which reduces the one-way ground-to-satellite propagation delays to a few milliseconds. As a consequence, the \gls{e2e} latency with \gls{ngso} satellites over long distances may be competitive and even lower than that of terrestrial networks~\cite{Handley2018}.
    \item \textbf{Global connectivity:} \Gls{ngso} satellites can provide coverage in remote areas where terrestrial infrastructure is not available. Furthermore, if appropriate functionalities are implemented, the data could be routed \gls{e2e} by the satellites themselves.
    \item \textbf{Feasible uplink communication from small devices:} Due to the relatively low altitude of deployment and the signals propagating mainly through free-space, it is feasible for small devices to communicate directly with \gls{leo} satellites. This has lead to companies and organisations to aim for integrated space and terrestrial infrastructures using \gls{lpwan} technologies such as LoRaWAN and \gls{nbiot} \cite{3GPPTR38821, Guidotti2019}.
    %\item \textbf{Independence:} The implementation details of the mechanisms of \gls{ngso} constellation are controlled by the designer and
\end{enumerate}

Based on these advantages, some of the main use cases for \gls{ngso} constellations include:
\begin{enumerate}
    \item \textbf{Backhauling:} Inter-satellite communication can be used to transmit the data towards the Earth, even when the source and destination are not within the coverage of the same satellite~  \cite{soret2020backhauling}.  
    \item \textbf{Offloading:} \Gls{ngso} constellations can serve as additional infrastructure in urban hot-spots where the capacity of the terrestrial network is temporarily exceeded, for example, during sport and cultural events.
    \item \textbf{Resilience:} Satellites in \gls{ngso} can serve as failback backhaul network for terrestrial \glspl{bs} in case the primary backhaul fails, e.g., due to natural disasters.
    \item \textbf{Edge computing and \gls{aiaas}:} \Gls{iot} devices have limited processing capabilities and limited energy supply (i.e., batteries). Therefore, \gls{ngso} satellites can be used as edge computing nodes~\cite{Xie2020} to reduce the computational load at the \gls{iot} devices. Furthermore, the satellites can gather data from several devices and locations along their orbit and use it, along with their computational capabilities, to provide \gls{aiaas} to devices where the data and/or processing capabilities are insufficient for \gls{ai}~\cite{Razmi2021}.
    \item \textbf{Earth observation:} \gls{ngso} satellites can be used as moving sensing devices that capture data, e.g., in the form of images or video, of physical phenomena at the Earth's surface or within its atmosphere.
    To obtain a sufficient resolution, \glspl{leo} are the preferred choice in most Earth observation satellite missions. Furthermore, sun-synchronous orbits, i.e., an orbit where the satellite maintains a constant angle towards the sun when viewed from Earth, often have favorable properties for Earth observation tasks. %These can be, e.g., nearly constant surface illumination angle, optimal solar panel placement towards the Sun, or using the Earth to block solar radiation to protect sensitive measurements. If precise station keeping is required, the trade-off between a lower orbit for better resolution and increased fuel consumption due to higher atmospheric drag needs to be taken into account. In some cases, a frozen orbit \cite{Capderou2014} might help to reduce the station keeping costs.
    %Another important aspect is that coordinated observation with multiple satellites can significantly improve the resolution, as exploited in existing missions, e.g., GRACE or TanDEM-X. In order to keep the delay between collecting the observed data and transmitting them to the ground, \glspl{isl} to satellites in higher orbits can be used.}
    %Usually these satellites are quite small and thus the energy requirements for station keeping should be as small as possible. Sun-synchronous orbits are therefore often preferred for Earth observation satellites. In contrast to most communication satellite they only communicate with dedicated ground stations. In order to keep the delay between collecting the observed data and transmitting them to the ground, \glspl{isl} to satellites in higher orbits are often used.}
\end{enumerate}

For the use cases mentioned above, and many more, \emph{global connectivity} is essential, as it allows to fully exploit the benefits of \gls{ngso} constellations. Specifically, it would allow the constellation to deliver the data generated on the ground, by aerial vehicles, or by the satellites themselves to the destination without heavily relying in additional (e.g., terrestrial) infrastructure. %Building on this, we classify the paths that the data can take in the constellation into the following four \emph{logical links}~\cite{Leyva-Mayorga2020}.
%\begin{description}
%\item{Ground to ground [G2G]:} With both, source and destination being ground and/or aerial terminals. This is the typical use of the constellation for terrestrial backhauling.

%\item{Ground to satellite [G2S]:} With the source being a ground or aerial terminal and the destination being a satellite. This link is mainly used for operations initiated by dedicated ground stations such as constellation, route, and link establishment and maintenance, tele-control and tele-command, and content caching.

%\item{Satellite to Ground [S2G]:} With the source being a satellite and the destination being a ground or aerial terminal. This link is mainly used when the satellites themselves generate application data, for example, in Earth and space observation, but also for telemetry, handover, link maintenance and adaptation, and fault reporting.

%\item{Satellite to Satellite [S2S]:} With the source and destination being satellites, possibly deployed at different altitudes and/or orbits. This link is used for applications including distributed processing and inference, sensing, but also for \gls{isl} establishment and adaptation, neighbour discovery, routing 
%\end{description}

Nevertheless, there are several \glspl{kpi} that should be considered to determine whether an \gls{ngso} constellation design is appropriate for the target application. These include, but are not restricted to
\begin{itemize}
    \item Service availability: The fraction of the time in which the ground terminal is able to communicate with the constellation~\cite{3GPPTR38821}. Through this chapter, we will assess this \gls{kpi} based on the coverage of the constellation in different locations, including remote (e.g., polar) regions. 
    \item Transport capacity: The maximum amount of data that can be transmitted by the constellation, \gls{e2e}, per time unit. 
    \item Throughput: Data rate experienced by the users.
    \item Scalability: Maximum number of devices supported by the constellation per unit area.
    \item Inter-satellite connectivity: The ability to achieve inter-satellite communication. Oftentimes it is assessed by the number of satellites with active connections~\cite{Leyva-Mayorga2021} or by the number of satellites within communication range~\cite{Kak2019}.  
    \item Latency and reliability: Probability that the data can be transmitted to the destination within a given time $t$.
    \item Energy efficiency: Since \gls{iot} devices and satellites are usually powered by batteries, minimizing the energy required for communication is essential. 
\end{itemize} 
Several of these \glspl{kpi} were considered by Del Portillo~\cite{DelPortillo2019} to compare the OneWeb, Starlink (outdated configuration with $h>1000$\,km), and Telesat constellations. 

Other \glspl{kpi} have been defined for satellite constellations. For example, Soret et al.~\cite{soret2020backhauling} emphasised the relevance of timing metrics beyond the packet delay, such as the \gls{aoi} and its by-products, for some satellite tracking or remote sensing applications.

%These links must be realised with a moving infrastructure, satellites moving w.r.t. each other and also w.r.t. the Earth through the satellite pass and due to the Earth's rotation~\cite{Ye2021}.

%\bs{based on all of this, there is a need to carefully design the constellation topology and capabilities: different services / applications / cost-performance-energy tradeoffs / small-big satellite technology}

\section{NGSO constellation design}
Satellite constellations are groups of satellites organized in orbital planes. The $N_\text{op}$ satellites in one orbital plane follow the same orbital trajectory, one after the other, and are usually uniformly spaced around the orbit. Furthermore, an orbital shell is a group of $P$ orbital planes in a constellation that are deployed at approximately the same altitude; some orbital shells may implement minor variations of a few kilometers called orbital separation. 
To maximize the coverage for communications, the organisation of satellites in one orbital shell usually belongs to one of two basic types: Walker star and Walker delta (also called Rosette)~\cite{Su2019,Qu2017, Walker1971, Walker1984}. Satellite constellation design may include one or more orbital shells.  
 
Walker star orbital shells consist of nearly-polar orbits, with typical inclinations of $\delta \approx90^\circ$, which are evenly spaced within $180^\circ$. As such, the angle between neighbouring orbital planes is $180/P$.   

Walker delta orbital shells, on the other hand, typically consist of inclined orbits, with typical inclinations of $\delta <60^\circ$, which are evenly spaced within $360^\circ$. As such, the angle between neighbouring orbital planes is $360/P$.

Due to the use of inclined orbits, Walker delta orbital shells do not provide coverage in polar regions nor in the northernmost countries such as Greenland. However, this allows to keep the satellites within the areas where most of the population resides and, hence, where data traffic is generated and consumed.

Since both Walker star and delta geometries provide distinct advantages and disadvantages, some companies such as SpaceX have considered a mixed design consisting of multiple orbital shells. Specifically, the design of the Starlink constellation considered a Walker delta orbital shell at around $550$\,km and at around $1100$\,km. However, the \gls{fcc} granted permission to SpaceX to modify the constellation geometry and to lower the $2814$ satellites in the $1100$\,km orbital shell to an altitude between $540$ and $570$\,km\,\cite{FCC2021}. Table~\ref{tab:constellations} shows the design parameters of some relevant \gls{ngso} constellations. The values in this table were obtained from the companies web pages, related papers~\cite{DelPortillo2019, Kak2019}, and \gls{fcc} filings and some of them have not yet been approved.\footnote{Updates on ongoing launches can be found at the New Space webpage \href{https://www.newspace.im/}{https://www.newspace.im/}}

Beyond the technical aspects, the dramatic increase in the number of objects put into orbit around the Earth due to the deployment of \gls{ngso} constellations has raised concerns on their long-term sustainability. Naturally, the more satellites orbit the Earth, the higher the risk of collision. Hence, measures to minimize the collision risk have been explored and should be adopted in commercial constellations~\cite{Lewis2019}. In particular, deploying the orbital planes at slightly different altitudes, with differences of less than $4$\,km, greatly reduces the collisions caused by failed satellites: a scenario that cannot be avoided. However, this introduces slight asymmetries in the constellations that complicate several technical aspects; these will be further described in Section~\ref{sec:link_establishment}. Another example of slight asymmetries in the constellations is that the satellites between neighbouring orbital planes may be shifted across the orbit, so that the satellites in one orbital plane are rotated by a relatively small angle w.r.t. those in the neighbouring planes.

\begin{figure}[t]
    \centering
    \includegraphics{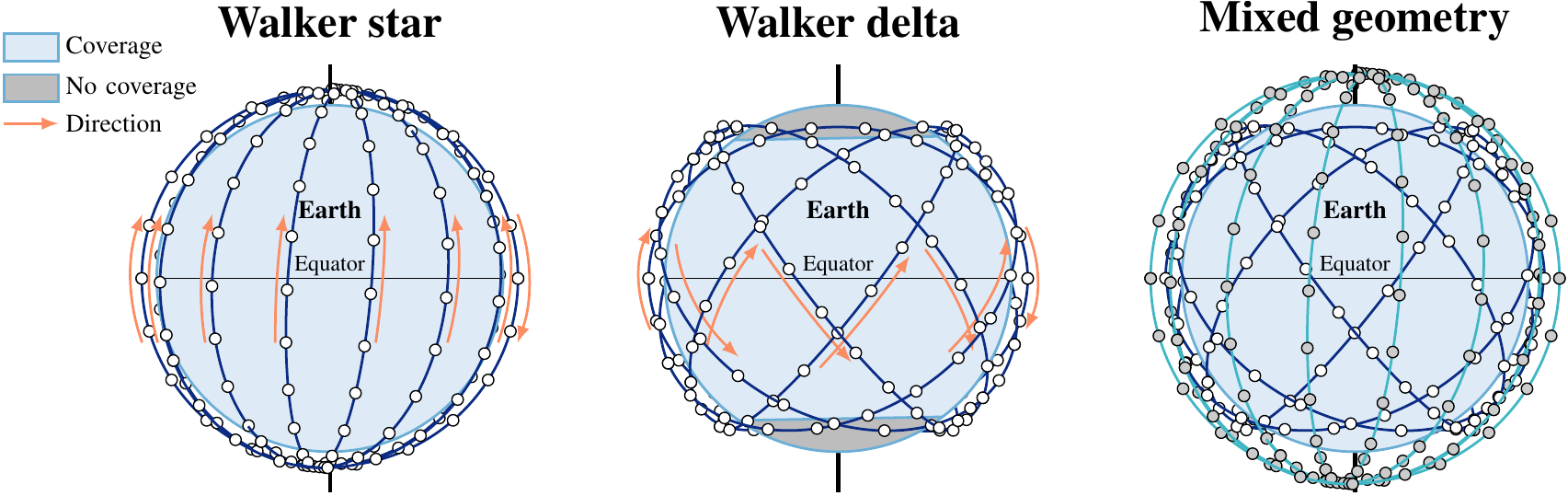}
    \caption{Diagram of Walker star, Walker delta (Rosette), and mixed constellation geometries.}
    \label{fig:constellation_geometries}
\end{figure}

\begin{table}[t]
    \centering
    \renewcommand{\arraystretch}{1.3}
    \footnotesize
    \caption{Parameters for some commercial \gls{ngso} satellite constellations}
    \begin{tabular}{@{}lllllllll@{}}
    \toprule
    Parameter & \multicolumn{7}{c}{Constellation}\\\cline{2-8}
    & \multicolumn{5}{c}{Starlink} & OneWeb & Kepler\\\midrule
    Type & \multicolumn{5}{c}{Mixed} & Walker star & Walker star\\\cline{2-6}
    Number of satellites $N$ & $1584$ &$1584$ & $720$ & $348$& $172$ & $648$ & $140$\\
    Number of orbital planes $P$&  $72$ & $72$& $36$ & $6$ &$4$ &$18$ & $7$\\
    Altitude $h$ (km) & $550$& $540$ & $570$ &$560$ & $560$ &$1200$ & $575$\\
    Inclination $\delta$ ($^\circ$) &$53$ &$53.2$ &$70$ &$97.6$& $97.6$& $86.4$ & $98.6$\\
    Intended service & \multicolumn{5}{c}{Broadband} & Broadband & \gls{iot}\\\bottomrule
    \end{tabular}
    
    \label{tab:constellations}
\end{table} 
\section{Communication links}
The endpoints for communication in an \gls{ngso} constellation can be either the ground or at satellite level. Therefore, the paths that the data can take in the constellation can be classified into the following four \emph{logical links}~\cite{Leyva-Mayorga2020}.
\begin{itemize}
\item{Ground to ground [G2G]:} With both, source and destination, being ground and/or aerial terminals. This is the typical use of the constellation for terrestrial backhauling.

\item{Ground to satellite [G2S]:} With the source being a ground or aerial terminal and the destination being a satellite. This link is mainly used for operations initiated by dedicated ground stations such as constellation, route, and link establishment and maintenance, tele-control and tele-command, and content caching.

\item{Satellite to Ground [S2G]:} With the source being a satellite and the destination being a ground or aerial terminal. This link is mainly used when the satellites themselves generate application data that must be transmitted to a ground station for storage and/or processing. For example, in Earth and space observation, but also for telemetry, handover, link maintenance and adaptation, and fault reporting.

\item{Satellite to Satellite [S2S]:} With the source and destination being satellites, possibly deployed at different altitudes and/or orbits. This link is used for localised network maintenance, updating routing tables, neighbour discovery, or other applications such as distributed processing, sensing and inference.
\end{itemize}

These links must be realised with a moving infrastructure. \gls{ngso} satellites move rapidly with respect to each other in higher and lower orbits and in different orbital planes.%\bho{NOTE: relative to each other, they don't really move when in same orbit}. 
They also move with respect 
to the Earth due to the satellite orbital velocity and to the Earth's rotation~\cite{Ye2021}. Specifically, the orbital velocity of the satellites $v_o$ is determined by the altitude of deployment $h$ as
\begin{equation}
    v_o(h)\approx \sqrt{\frac{\text{G}M_E}{R_E+h}},
    \label{eq:orb_vel}
\end{equation}
where $\text{G}$ is the universal gravitational constant; $M_E$ and $R_E$ are the mass and radius of the Earth, respectively.
\rev{Then, according to Kepler's third law of planetary motion, the orbital period of a satellite can be closely approximated as
\begin{equation}
    T_o(h)\approx \frac{2\pi(R_E+h)}{v_o(h)}=\sqrt{\left(\frac{4\pi^2}{\text{G}M_E}\right)\left(R_E+h\right)^3}.
    \label{eq:orb_period}
\end{equation}}
From here, and assuming traditional \gls{leo} altitudes, for example, with $h=600$\,km, it is easy to observe that the orbital velocity of \gls{ngso} satellites may exceed $7.6$\,km/s and that their orbital period is usually around $90$ minutes. 

Furthermore, satellites in different locations of the constellation may move rapidly w.r.t. to each other. Finally, the whole satellite constellation is moving w.r.t. the Earth due to its rotation~\cite{Ye2021, Qu2017}. Hence, an important aspect to select the altitude of deployment of a constellation is whether it is desired that the orbit is \emph{recursive}. That is, whether  the satellites should pass over the same point in the Earth at a specific time of the day after a given number of days $m$. For this, we require to find an altitude $h_\text{rec}$ for which $n T_o = m T_E$, where $T_E=86164$\,s is the equinoctial day~\cite{Qu2017}. \rev{To find the required altitude for the recursive orbit, let us first rewrite the right-hand side of~\ref{eq:orb_period} as
\begin{equation}
    T_o^2=\left(\frac{4\pi^2}{\text{G}M_E}\right)\left(R_E+h\right)^3,
\end{equation}
which allows us to define $h$ as a function of $T_o$ as
\begin{equation}
    h=\left(\frac{T_o^{2}\text{G}M_E}{4(\pi)^2}\right)^{1/3}-R_E.
    \label{eq:alt_to_period}
\end{equation}
Finally, we substitute the period $T_o=mT_E/n$ in~\eqref{eq:alt_to_period} to find the altitude for a recursive satellite orbit as
\begin{equation}
    h_\text{rec}(n,m,T_E)=\left(\frac{(mT_E)^{2}\text{G}M_E}{(2n\pi)^2}\right)^{1/3}-R_E
    \label{eq:rec_period}
\end{equation}}
From~\eqref{eq:rec_period} we obtain that \gls{ngso} satellites at $h=554$\,km, close to Starlink's altitude of deployment, have recursive orbits for $n=15$ and $m=1$. That is, these will orbit the Earth exactly $15$ times each day. Moreover, those at $h=1248$\,km, close to OneWeb's altitude of deployment, have recursive orbits for $n=13$ and $m=1$.

An essential aspect to observe about \gls{ngso} satellite constellations is that, even though the relative positions and velocities of satellites w.r.t. other satellites and to the ground terminals are dynamic, the dynamics of the constellation are fully dictated by the physics of the system and, hence, completely predictable. Therefore, the topology of the network (space and terrestrial) at a point in time $t$ can be perfectly predicted with a high level of certainty. Because of this, approaches from ad-hoc networks~\cite{Marcano2020} as well as from fully structured networks have been applied in the context of \gls{ngso} satellite constellations. 

Furthermore, the different time scales of the various ongoing processes offer opportunities for simplification via time-scale separation. For instance, the orbital period of a satellite is extremely long when compared to most of the communication tasks within the constellation. Therefore, the satellite constellation can be assumed to be static during short periods to simplify the analysis. In the following, we exemplify this later aspect by calculating one-hop latency of the different links.

Depending on whether we consider a \gls{gsl} or an \gls{isl}, the one-hop latency is determined by different factors. Naturally, it depends on the position of the transmitter $u$ and the receiver $v$ at time $t$, when the packet is ready to be transmitted and also on the packet length $p$. In the following, we calculate the three main components of the one-hop latency by considering the relative position of $u$ w.r.t. $v$ to be fixed during a period $[t, \Delta t]$.

First, the \textit{waiting time} at the transmission queue $q_t(u,v)$ is the time elapsed since the packet is ready to be transmitted until the beginning of its transmission. Note that, depending on the communication protocols, for example, signaling, and frame structure, it may occur that $q_t(u,v)>0$ for all $u,v$ even when there are no more packets in the queue. Second, the \textit{transmission time}, which is the time it takes to transmit $p$~bits at the selected rate $R_t(u,v)$~bps. Third, the \textit{propagation time}, which is the time it takes for the electromagnetic radiation to travel the distance $d_t(u,v)$ from $u$ to $v$. Hence, the latency to transmit a packet of size $p$ from $u$ to $v$ at time $t$ is given by
\begin{equation} 
    L_t(u,v) = \!\underbrace{ q_t(u,v) }_\text{Waiting time}+ \!\underbrace{ \dfrac{ p }{ R_t(u,v) } }_\text{Transmission time}+\underbrace{ \dfrac{d_t(u,v)}{c} }_\text{Propagation time}.
    \label{eq:packet_latency}
\end{equation}

Note that all the factors that contribute to the one-hop packet latency depend on the time the packet is generated. Furthermore, due to the movement of the satellites, the set of established links and communication paths (routes) change depending on $t$.
This creates a greatly dynamic network topology that introduces distinctive challenges in the design and implementation of the distinct physical links. In the following, we elaborate on the main technologies for satellite communications: \gls{rf} and \gls{fso} links.

\emph{\Gls{rf} links} occur in frequencies either in the S-band, the Ka-band or the Ku-band. These links are mainly affected by free-space path loss and thermal noise, so \gls{awgn} channels are oftentimes considered. The free-space path loss between two terminals $u$ and $v$ at time $t$ is determined by the distance $d_t(u,v)$ between them and the carrier frequency $f$ as 
\begin{equation}
\mathcal{L}_t(u,v)=\left(\frac{4\pi d_t(u,v) f}{c}\right)^2,    
\end{equation}
where $c$ is the speed of light.

Next, let $P^{(u)}$ be the transmission power of transmitter $u$ -- assumed to be constant for simplicity -- and $\sigma^2_v$ be the noise power at receiver $v$. Further, let $G_t^{(u,v)}$ and $G_t^{(v,u)}$ be the antenna gain of transmitter $u$ towards receiver $v$ and vice versa. 
Based on this, the maximum data rate for reliable communication between two satellites and/or a satellite and a ground terminal at time $t$ can be calculated as
a function of the \gls{snr}
\begin{equation}
R_t(u,v) =B\log_2\Big(1+\text{SNR}_t\left(u,v\right)\Big) =B\log_2\left(1+ \frac{P^{(u)}G_t^{(u,v)}G_t^{(v,u)}}{\mathcal{L}_t(u,v)\sigma^2_v}\right).
\label{eq:rate}
\end{equation}
Naturally, the achievable data rate in the presence of interference will be lower than \eqref{eq:rate}. Nevertheless, the use of directional antennas and/or orthogonal resource allocation~\cite{Leyva-Mayorga2021} greatly reduces interference within constellations. Building on this, the achievable rate mainly depends on the transmission power, the large-scale fading (path loss), and noise power, but also on the gain of the communicating antennas in the direction of the receiver/transmitter. Since the constellation is a moving infrastructure, antenna pointing technology is an essential aspect of constellation design. 

Throughout this chapter, we evaluate the performance of the \gls{rf} physical links by assuming the parameters listed in Table~\ref{tab:comm_params} unless stated otherwise. These parameters were selected to focus on comparing the constellation design and not the implemented (envisioned) communication technologies.

\begin{table}[t]
    \centering
    \footnotesize
     \renewcommand{\arraystretch}{1.3}
    \caption{Parameter configuration for the physical links: \gls{gsl} and \gls{isl}.}
    \begin{tabular}{@{}llll@{}}
         \hline
         Parameter & Symbol & NGEO to GS
         &ISL  \\\hline
         Carrier frequency (GHz)& $f$ & $20$ & $26$\\
         Bandwidth (MHz) & $B$ & $500$ & $500$\\
         Transmission power (W) & $P_t$ & $10$ & $10$\\
         Noise temperature (K) & $T_N$ & $150$ & $290$\\
         Noise figure (dB) & $N_f$ & $1.2$ & $2$\\
         Noise power (dBW)& $\sigma^2$ &$-117.77$ & $-114.99$\\
         \textbf{Parabolic antennas}\\
         \quad Antenna diameter (Tx -- Rx) (m)& $D$ & ($0.26$ -- $0.33$) & ($0.26$ -- $0.26$)\\
         \quad Antenna gain (Tx -- Rx) (dBi) & $G_\text{max}$  & ($32.13$ -- $34.20$) & ($34.41$ -- $34.41$)\\
         \quad Pointing loss  (dB) & $L_p$ &  $0.3$ & $0.3$\\
         \quad Antenna efficiency (--) & $\eta$ & $0.55$ &  $0.55$ \\
         
         \hline
    \end{tabular}
    \label{tab:comm_params}
\end{table}

\emph{\Gls{fso} links}, on the other hand, face different challenges depending on where they are implemented: \gls{gsl} or \gls{isl}~\cite{Kaushal2017}. Hence, these challenges will be briefly described in the following sections.

\subsection{Ground-to-satellite links (GSLs)}
Communication between devices deployed at ground level and the satellites takes place through \glspl{gsl}. This can occur either by communicating the user devices (e.g., \gls{iot} devices) directly or through gateways. The gateways can be deployed at ground level, but also in the air, such as  \glspl{uav} or \glspl{hap}. For simplicity, we use the term ground terminal to refer to any device deployed at ground level. The area where ground-to-satellite communication is possible is called the \emph{coverage area} and the time the satellite and a terrestrial terminal can communicate is called the duration of the \emph{satellite pass}. 

In the following, we provide the expressions to calculate the coverage area and, hence, to determine whether a ground terminal is able to communicate with a specific satellite at a given time $t$. 

 The distance between an \gls{ngso} satellite and a device located on the Earth's surface within line-of-sight at time $t$ is determined by the altitude $h$ and the elevation angle of the satellite w.r.t. the device $\varepsilon_t$. \rev{Specifically, the distance of the \gls{gsl} can be calculated from these parameters using the Pythagorean theorem in a triangle with sides of length: a) $R_E+h$; b) $R_E+d_\text{GSL}(h, \varepsilon_t)\sin \varepsilon_t$; and c) $d_\text{GSL}(h, \varepsilon_t)\cos \varepsilon_t$ and then applying the quadratic formula to obtain:
\begin{equation}
    d_\text{GSL}(h, \varepsilon_t) = \sqrt{R_E^2\sin^2(\varepsilon_t)+2R_Eh+h^2}-R_E\sin(\varepsilon_t).
\end{equation}
A similar procedure can be applied to the case of devices above the Earth's surface (e.g., \glspl{uav}, \glspl{hap}, etc.), by simply substituting the length of side b) of the triangle to be $R_E+h_u+d_\text{GSL}(h, \varepsilon_t)\sin \varepsilon_t$, where $h_u$ is the altitude of the user above the sea level $R_E$. For notation simplicity, the rest of the equations presented are for satellites deployed at the Earth's surface only. However, a the substitution described above can be used to adapt the following equations to devices deployed above the Earth's surface.}

\rev{Once $d_\text{GSL}(h, \varepsilon_t)$ has been found, we calculate the Earth central angle~\cite{Kak2019}  $\alpha(h, \varepsilon_t)$ as}
\begin{equation}
    \alpha(h, \varepsilon_t)=\arccos\left(\frac{(R_E+h)^2+R_E^2-d_\text{GSL}(h, \varepsilon_t)^2}{2(R_E^2+hR_E)}\right),
\end{equation}
 which determines the shift in the position of the device w.r.t. the satellite's nadir point.

The coverage of an \gls{ngso} satellite is usually defined by a minimum elevation angle $\varepsilon_\text{min}$. Hence, a device located at an elevation angle $\varepsilon_t\geq \varepsilon_\text{min}$ is considered to be within coverage of the satellite at time $t$.
Consequently, the coverage area of a satellite is a function of the altitude of deployment $h$ and of $\varepsilon_\text{min}$. By using $h$ and $\varepsilon_\text{min}$ we find the angle $\alpha(h,\varepsilon_\text{min})$, which allows us to calculate the coverage area as
\begin{equation}
    A(h, \varepsilon_\text{min})=2\pi R_E^2(1-\cos(\alpha(h, \varepsilon_\text{min})).
\end{equation}
Furthermore, by assuming a spherical model of the Earth, we can easily determine whether a ground terminal $u$ is within coverage of a satellite $v$ at a given time $t$; this occurs when the distance $d_t(u,v)$ between them is shorter than  $d_\text{GSL}(h,\varepsilon_\text{min})$. 

Next, we calculate the maximum duration of a satellite pass as a function of $\alpha(h, \varepsilon_\text{min})$ and $T_o(h)$. For this, let $t=0$ be the time when the ground terminal enters the coverage area of the satellite. The satellite pass has maximum duration in the case where, at exactly at the middle of the pass, the satellite is exactly located at the zenith point of the ground terminal and, hence, there is an angle $\epsilon_t=90^\circ$ between the terminal and the satellite w.r.t. the Earth's center. 
In such case, the satellite travels $\alpha(h,\varepsilon_\text{min})/180$\,degrees of its orbit and hence, the satellite pass has a duration
\begin{equation}
    T_\text{pass}(h,\varepsilon_\text{min})\leq \frac{T_o(h)\alpha(h,\varepsilon_\text{min})}{\pi}.
\end{equation}
For any other cases where the ground terminal and the satellite are not perfectly aligned, we define the angle
\begin{equation}
    \alpha_\text{min}=\min_{t} \alpha(h,\varepsilon_t) \quad \text{s.t. } t\in\left[0, T_\text{pass}(h,\varepsilon_\text{min})\right],
\end{equation} 
which determines the misalignment of the ground station w.r.t. the orbital plane of the satellite. Naturally, $\alpha_\text{min}=0$ for the perfect alignment case.

\begin{figure}[t]
    \centering
    \subfloat[]{\includegraphics{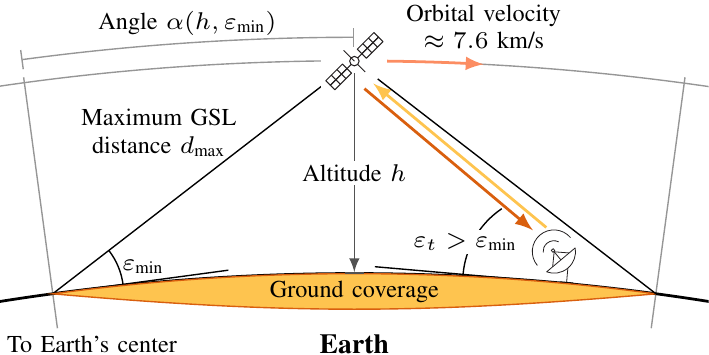}}\\
    \subfloat[]{
\includegraphics{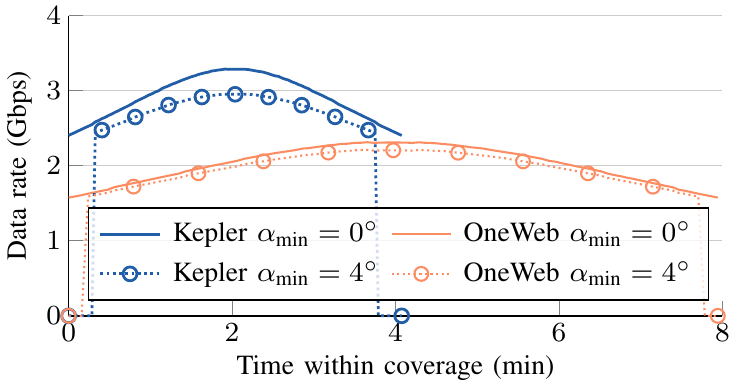}
\label{fig:Rate_pass}}
    \caption{(a) Ground coverage of an \gls{ngso} satellite at altitude $h$ and (b) the evolution of the achievable data rate along the pass.}
    \label{fig:gsl}
\end{figure}

The ground coverage of a \gls{ngso} satellite is illustrated in Fig.~\ref{fig:gsl} and the evolution of the achievable data rate along the pass for the altitudes of deployment for Kepler and OneWeb. \rev{We considered a typical value for the minimum elevation angle of $\varepsilon_\text{min}=30^\circ$. For devices deployed above the Earth's surface, this angle may be larger as their \gls{los} is less affected by obstacles}. From these, it is easy to observe that lower altitudes of deployment result in shorter propagation delays but also in faster orbital velocities, shorter satellite passes, and smaller coverage areas.

Note that the coverage area simply defines the area where communication is possible. However, the beams oftentimes present a beamwidth that is much narrower than the coverage area. Therefore, these must be pointed in the desired direction of communication~\cite{3GPPTR38821}. Because of this, having more than one satellite within communication range can be beneficial as the access load can be shared among the satellites covering the same areas. Hence, it provides an indicator of the scalability and capacity of the network.

\begin{figure}[t]
    \centering
    \subfloat[]{\includegraphics{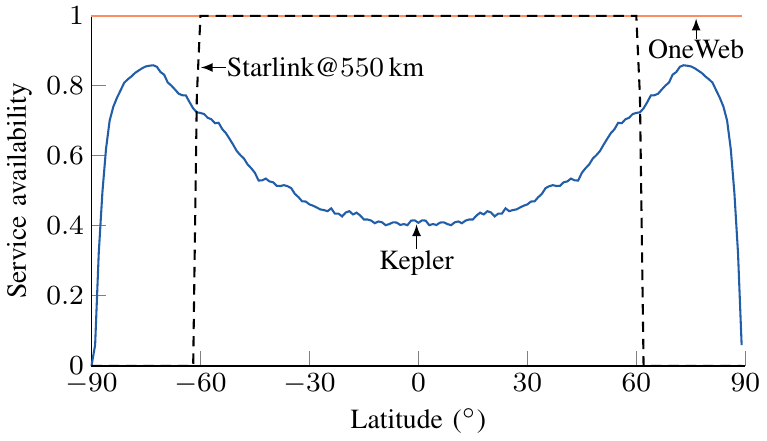}\label{fig:availability}}\hfil
    \subfloat[]{\includegraphics{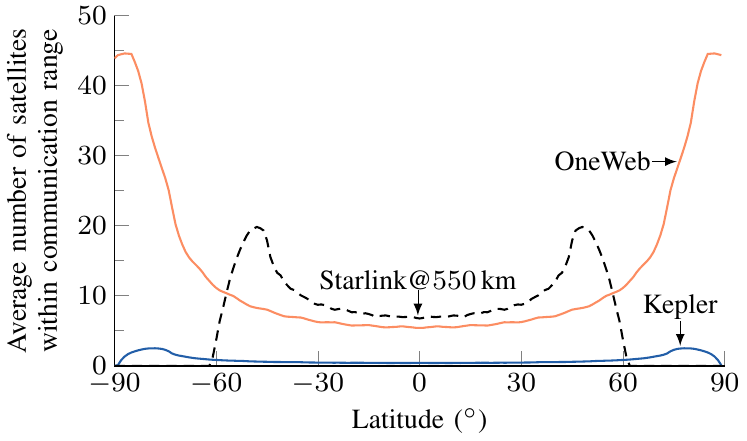}\label{fig:coverage}}

    \caption{(a) Service availability: probability of being within the coverage area of a satellite as a function of the latitude and (b) average number of satellites within communication range at \gls{gsl}.}
    \label{fig:coverage_availability}
\end{figure}

Based on the coverage area and the geometry of a specific constellation, the service availability and the average number of satellites within range can be obtained.  
Fig.~\ref{fig:coverage_availability} shows the service availability and the mean number of satellites within coverage for the Kepler and OneWeb constellations, along with the Starlink orbital shell at $h=550$\,km considering the requested modification in the latest \gls{fcc} filing, where $\varepsilon_\text{min}=25^\circ$. 

As it can be seen, the density of the Kepler constellation and the considered $\varepsilon_\text{min}=30^\circ$ are insufficient to provide full service availability near the Equator and it increases in near-polar areas. In contrast, the service availability of the Starlink orbital shell is guaranteed between latitudes $[-60^\circ,60^\circ]$ and the OneWeb constellation provides full service availability across the globe. Furthermore, it can be seen in Fig.~\ref{fig:coverage} that there is a significant number of OneWeb satellites within coverage in the polar regions and a considerably lower number in Equatorial regions. This is a distinctive characteristic of Walker star constellations, as the distances between satellites is maximal near the Equator. In contrast, the coverage of the Starlink orbital shell between the latitudes $[-60^\circ,60^\circ]$ is relatively balanced. To solve the problem of lack of coverage in the polar regions, the Starlink constellation is planned to incorporate satellites in polar orbits, as listed in Table~\ref{tab:constellations}.

There are many benefits of using \gls{rf} over \gls{fso} for the \gls{gsl}. For instance, \gls{rf} links present a wider beamwidth and, hence, a broader coverage. This simplifies the beam switching and allows to provide coverage to several ground terminals simultaneously. Additionally, the use of \gls{rf} links allows to use the same physical layer technologies as in terrestrial networks, which simplifies the hardware design and enables the integration of satellites and terrestrial networks through mature terrestrial technologies. For instance, the \gls{3gpp} is aiming to integrate satellites and cellular networks using \gls{nbiot} and 5G \gls{nr} cellular technologies~\cite{3GPPTR38821, 3GPPTR38913, Liberg2020}. 

In contrast, \gls{fso} \gls{gsl} are mainly affected by the atmosphere. In particular, the atmosphere absorbs and scatters the beam. These effects depend on different factors such as temperature, humidity, and the concentration of aerosol particles. Furthermore, the effects vary widely between uplink and downlink communication, with the uplink signals being affected most due to the presence of the atmosphere around the transmitter~\cite{Kaushal2017}.

Yet another factor that impacts the \gls{gsl} is the Doppler shift. The latter varies significantly during a satellite pass as a result of the high orbital velocity in combination with the varying relative position and speed with respect to time. That is, the Doppler shift is different between the edge and the center of the coverage, so this must be taken into account to select an appropriate frequency band and during waveform and antenna design. If information from the Global Navigation Satellite System (GNSS) is available, the Doppler shift can be first pre-compensated at the satellite w.r.t. to a reference point, by exploiting the predictable movement of the satellites. Then, the residual frequency offsets are compensated at the ground terminals using traditional Doppler compensation techniques as in terrestrial networks~\cite{3GPPTR38821}. %\bho{Reference?}\mr{Note: Moved the 3GPP reference to end of the section,  because compensation aspects are discussed there as well.}

%Modulation schemes for satellite communications usually involve low modulation orders for robustness and low peak-to-average power ratio (PAPR) to overcome the high path loss by driving the power amplifiers into saturation.
%enable the use of nonlinear power amplifiers.
%The preferred choice in recent commercial LEO missions is amplitude and phase-shift keying (APSK) with modulation orders up to $16$~\cite{DelPortillo2019}. Hence, the most suitable modulation schemes supported by 5G NR are quadrature phase-shift keying (QPSK) and quadrature amplitude modulation (QAM) with modulation order $16$.
%Terrestrial gNBs adapt the modulation and coding scheme to the current channel conditions, for which the UEs must transmit information about the channel quality to the gNB~\cite{Guidotti2019}.

%\mr{For the transmission of short packets in the GSL, the priority is to minimize link outages and packet errors to avoid long RTTs for feedback. In these cases, fixed robust modulation and coding schemes are preferred.  On the other hand, adaptive coding and modulation is interesting for long packet transmissions in the GSL, where the predictability of the path loss can be exploited.}

%%%%%%%%%%%%%%%%%%%%%%%%%%%%%%%%%%%%%%%%%%
\subsection{Inter-satellite links (ISLs)}
Inter-satellite communication takes place
%The \glspl{isl} communicate satellites
in 1) the same orbital plane, 2) different orbital planes of the same orbital shell, and 3) different orbital altitudes. The dynamics in each of these are significantly different. However, it is essential to establish these links in an efficient manner to maximize the connectivity within the constellation.

Intra-plane \glspl{isl} connect satellites in the same orbital plane, usually, at both sides of the roll axis, which is aligned with the velocity vector. In particular, the relative distances between neighbouring satellites within the orbital plane -- the intra-plane distance -- at an altitude can be considered a constant 
\begin{equation}
    d_\text{intra}(N_\text{op},h)=2(R_E+h)\sin\left(\frac{\pi}{N_\text{op}}\right).
\end{equation}
Hence, intra-plane \glspl{isl} are rather stable. Still, the orbital velocity of the satellites must be considered. But this is easily, compensated by selecting an appropriate \gls{paa}: instead of pointing the antennas directly towards the instantaneous position of the receiver at the same time instant $t$, they are pointed to its position after considering the propagation time $t+d_\text{intra}(h)/c$.
Because of this, the antennas used for intra-plane communication can be highly directive and the beams can be fixed to the appropriate direction. Due to the use of narrow beams, \gls{fso} links present an interesting option for intra-plane communication, as their power efficiency may be greater than that of \gls{rf} links~\cite{Kaushal2017}. Nevertheless, \gls{rf} links with either parabolic or patch antenna arrays are also an efficient candidate that combines relatively high gains, cheap components, and low power requirements when compared to \gls{fso}.

Inter-plane \glspl{isl}, on the other hand, connect satellites within the same orbital shell but in different orbital planes. Usually, satellites will possess either one or two transceivers for inter-plane communication, with antennas pointing towards both sides of the pitch axis.  Depending on the constellation geometry, the distances and the velocity vectors between satellites in different orbital planes may be either very similar or vary widely. For instance, in Walker star geometries, the orbital planes are separated by the angle $\pi/P$ and the shortest inter-plane distances occur at the crossing points of the orbits near the poles. In contrast, the longest inter-plane distances to the nearest neighbour occur for satellites near the Equator. 

Let $u$ and $v$ be a pair of satellites located in neighbouring orbital planes, where $v$ is the closest inter-plane neighbour of $u$ at time $t$. For simplicity, we assume the same altitude of deployment for both orbital planes to be $h$. We denote the polar angle of satellites $u$ and $v$ as $\theta^{(u)}_t$ and $\theta^{(v)}_t$, respectively. \rev{First, we recall that the distance between two points $u$ and $v$ on a sphere of radius $R_E+h$ with azimuth angles $\phi_u$ and $\phi_v$ is given as
\begin{IEEEeqnarray}{l}
d_{uv}(t)=\sqrt{2\left(\mathrm{R_E}+h\right)^2\left(1- \cos\theta_{t}^{(u)}\cos\theta_{t}^{(v)}-\cos\left(\phi_u-\phi_v\right)\sin\theta_{t}^{(u)}\sin\theta_{t}^{(v)}\right)}.\IEEEeqnarraynumspace
 \label{eq:distance_spherical}
\end{IEEEeqnarray}
The latter can be used to approximate the distance between two satellites adjacent orbital planes in a Walker star constellation assuming perfectly polar orbits. For this, recall that orbital planes in Walker star constellations are separated by $\pi/P$, hence, this is also the azimuth angle between satellites in adjacent orbital planes. }

\rev{If $u$ and $v$ are exactly at the Equator we have $\theta^{(u)}_t=\theta^{(v)}_t=\pi/2$, and the maximum intra-plane distance for the case where the satellites are perfectly aligned at all times only depends on $P$ and $h$ as
\begin{equation}
    d_\text{inter, aligned}^*(P)=\sqrt{2(R_E+h)^2\left(1-\cos\left(\frac{\pi}{P}\right)\right)}=2(R_E+h)\sin\left(\frac{\pi}{2P}\right).
\end{equation}
However, in a general case where the satellites $u$ and $v$ are not perfectly aligned, we have that, if $v$ is the closest inter-plane neighbour to $u$, then it follows that $|\theta^{(v)}_t -\theta^{(u)}_t|\in \left[0,\pi/N_\text{op}\right]$.
Therefore, the maximum inter-plane distance occurs when $\theta^{(u)}_t=\pi/2$ and $\theta^{(v)}_t = \pi/2\pm \pi/N_\text{op}$, which can be approximated as
\begin{IEEEeqnarray}{rCl}
d_\text{inter}^*(N_\text{op},P) &=& \max_{t}~ d_{uv}(t) \quad \text{s.t. } \theta^{(v)}_t\in\left[-\pi/N_\text{op},\pi/N_\text{op}\right] \IEEEnonumber\\ 
\quad&\approx&  \left(\mathrm{R_E}+h\right)\sqrt{ 2
-2 %\right.\nonumber\\ & \left.
\cos\left(\frac{\pi}{P}\right)\sin\left(\frac{\pi}{2}\pm\frac{\pi}{N_\text{op}}\right) }.\IEEEeqnarraynumspace
 \label{eq:maxdistance}
\end{IEEEeqnarray}}
Hence, to ensure that satellites at the Equator can communicate with at least one of their inter-plane neighbours, it is necessary to ensure that a non-zero data rate can be achieved at this location. To illustrate this aspect in a general case where the satellites are not perfectly aligned, let $\mathcal{R}$ be the set of available rates for communication, which depend on the available \glspl{mcs} and where $0\notin\mathcal{R}$. Then, to guarantee global \gls{isl} connectivity, it is required that any given satellite $u$ can select a rate $R\in \mathcal{R}$ that allows it to achieve reliable communication with the nearest inter-plane neighbour $v$ at all times. Hence, global \gls{isl} connectivity is achieved if
\begin{equation}
   \exists R\in\mathcal{R} : 0<R< B\log_2\left(1+ \frac{P^{(u)}G^{(u,v)}_tG^{(v,u)}_t c^2}{\left(2\pi\, \sigma_v \, d^*_\text{inter}(N,P) f\right)^2}\right).
\end{equation} 
As it can be seen, for a fixed set of rates $\mathcal{R}$, global \gls{isl} connectivity can be achieved by either increasing the power and/or gains of the antennas or by decreasing the maximum inter-plane distances. The latter is usually achieved by either increasing the number of orbital planes $P$, but also the number of satellites per orbital plane $N_\text{op}$. The interested reader is referred to our previous work for a general formulation that considers orbital separation and where the effect of increasing the number of orbital planes $P$ is illustrated~\cite{Leyva-Mayorga2021}.

Yet another characteristic of Walker star constellations is that the velocity vectors of the satellites in neighbouring orbital planes usually point in a similar direction. As a result of this, the relative velocities between these satellites are relatively low. However, there are specific pairs of orbital planes where the velocity vectors point in a nearly opposite direction: the so-called cross-seam \glspl{isl}. In the latter, the relative velocity of the satellites increases to nearly $2v_o$ and varies along with time.

\begin{figure}
    \centering
    \includegraphics{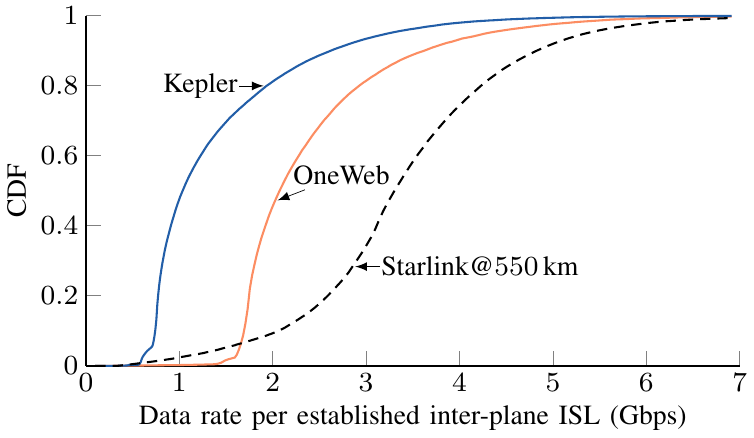}
    \caption{\gls{cdf} of the achievable data rate per inter-plane \gls{isl} with parabolic antennas.}
    \label{fig:rates_parabolic}
\end{figure}

As a consequence of these great differences, the Doppler shift and the \emph{contact times} in the inter-plane \gls{isl} --- the period where two satellites can communicate --- also vary widely. Therefore, it is essential to consider the movement of the satellites to select the inter-plane  \gls{isl} that must be established and to point the beams in the desired directions~\cite{Leyva-Mayorga2021, Leyva-Mayorga2021GC}. Fig.~\ref{fig:rates_parabolic} shows the \gls{cdf} of the achievable rate in the inter-plane \gls{isl} with a specific link establishment mechanism. While the mechanisms for \gls{isl} establishment and beam pointing are described in Section~\ref{sec:link_establishment}, Fig.~\ref{fig:rates_parabolic} shows that the data rate achieved by inter-plane \gls{isl} in the Starlink orbital shell is considerably larger than in the OneWeb and Kepler constellations. The main reason for this is the higher density of satellites caused by the low altitude of deployment, the use of Walker delta geometry, and, naturally, the large number of satellites.

Finally, inter-orbit \glspl{isl} connect satellites between different orbital altitudes~\cite{Kaushal2017}. For example, they can connect \gls{leo} satellites in different orbital shells or \gls{leo} satellites with \gls{meo} or even \gls{geo} satellites. A clear example are the \gls{fso} links in the \gls{edrs} and those envisioned to connect the different orbital shells in the Starlink constellation.

%The Doppler shift in the intra-plane \gls{isl} is constant and can be easily compensated during the design of the \gls{rf} links. For example, by simply implementing an appropriate offset in the carrier frequency between transmitter and receiver. 

\section{Functionalities and challenges}

\subsection{Physical Layer}
Pure \gls{los} connections, high velocities, and large transmission distances between satellites and ground terminals introduce some unique characteristics to the physical layer design for \gls{ngso} constellations, both in the \gls{gsl} and the \gls{isl}. 

In the \gls{gsl}, it is particularly appealing to maintain the waveforms used in terrestrial systems, for example, \gls{ofdm} in 5G \gls{nr} and \gls{nbiot} \cite{Kodheli2017,Guidotti2019}. This would allow full compatibility of terrestrial devices and direct \gls{iot}-to-satellite access which, in turn, grants maximum flexibility of deployment following the place-and-play vision. However, the subcarrier spacing in terrestrial \gls{ofdm} systems is narrow -- between $3.75$\,kHz for \gls{nbiot} and from $15$ to $240$\,kHz for 5G \gls{nr}~\cite{3GPP38211}. Such narrow subcarrier spacings make \gls{ofdm} highly sensitive to Doppler shifts and thus, accurate Doppler compensation is required to achieve reliable communication.
To overcome these limitations, several alternatives have been studied intensively in the literature over the past few years, such as Universal Filtered Multi-Carrier (UFMC), Generalised Frequency Division Multiplexing (GFDM) and Filter Bank Multi-Carrier (FBMC)~\cite{Wunder2014}. These waveforms allow for higher robustness against Doppler shifts and flexible time-frequency resource allocation in exchange for a higher equalisation complexity. However, in case of severe Doppler shifts, Factor Graph based equalisation for FBMC transmissions outperforms the OFDM system in terms of complexity and performance \cite{Woltering2018}. 

Another challenge for keeping reliable \gls{gsl} and also \gls{isl} is the implementation of adaptive modulation and coding. In \gls{3gpp} networks, the users exchange information about the channel quality with the \gls{bs}~\cite{Guidotti2019}, which adapts the \gls{mcs} based on the error rate. Due to the altitude of deployment, the round trip time between the ground terminals and a satellite is usually greater than $4$\,ms. Hence, such a feedback link would introduce a significant delay. Instead, the fully predictable movement of the satellite along the pass, in combination with free-space propagation and the minor impact of atmospheric conditions in \gls{rf} links, can be exploited to achieve efficient adaptive modulation and coding with minimal signaling. 

\begin{figure}[t]
\centering
\subfloat[]{
\includegraphics{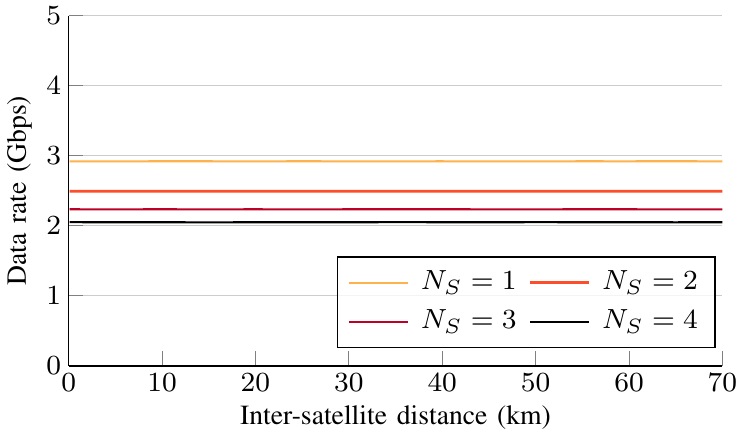}
\label{fig:Rate_swarms_Nr=1}}\\
\subfloat[]{\includegraphics{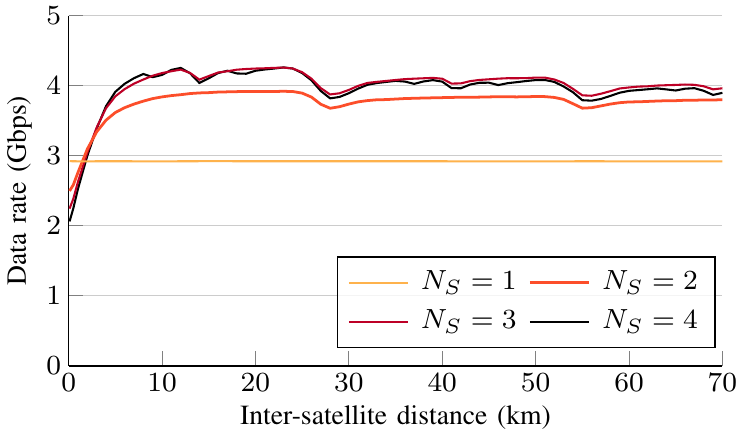}\label{fig:Rate_swarms_Nr=6}}
\caption{Data rate for satellite swarms as a function of the inter-satellite distance for (a) one and (b) six receiving antennas.}
\label{fig:Rate_swarms}
\end{figure}

Furthermore, while \gls{mimo} techniques have experienced a dramatic surge of advancements in terrestrial networks, achieving efficient \gls{mimo} communication with \gls{ngso} satellites is more complicated.
In particular, due to the long distances between transmitter and receiver, exploiting the full \gls{mimo} gain requires a large array aperture, that is, large distances between transmit and/or receive antennas, that are not feasible in individual satellites~\cite{Schwarz2019}. Nevertheless, this separation can be realised by using a group of satellites flying in close formation, usually called a \emph{satellite swarm}. Specifically, by placing an antenna at each of the satellites in the swarm, these can operate as distributed \gls{mimo} arrays. Doing so allows to form extremely narrow beams for \gls{gsl}, which leads to better spatial separation via coordinated beamforming and, eventually, to higher spectral efficiency when serving different ground terminals located geographically close to each other~\cite{Roper2019}. An example of the achievable gain of distributed \gls{mimo} in a satellite swarm, with $N_S$ satellites, is shown in Fig.~\ref{fig:Rate_swarms} for $N_r=1$ and $N_r=6$ receiving antennas. The overall transmit power as well as the antenna gains are normalised such that they are the same in all scenarios, i.e., the transmit power and antenna gain per satellite are $10/N_S$\,W and $32.13\,\text{dBi} - 10\log_{10}(N_S)$, respectively, and the receive antenna gain is $34.20\,\text{dBi} - 10\log_{10}(N_r)$.
Fig.~\ref{fig:Rate_swarms} shows that, despite maintaining the total transmitted power in all cases, the use of distributed \gls{mimo} increases the data rate by around $33$\% with multiple receiving antennas. However, with a single receiving antenna, no MIMO gains can be achieved and resulting in even lower rates because the transmitted signals superimpose constructively or destructively with same probability, reducing the overall received signal energy.

%Eventually, the use of large antenna arrays is one of the key technologies in NR, with promising application in GSLs as well. Such arrays allow to steer the beams very fast via digital processing techniques, without the need of mechanically steering the antenna. 

Beam pointing/steering is another essential functionality in \gls{ngso} constellations due to the constant and rapid movement of the satellites. Mechanical steering of \gls{rf} antennas becomes problematic as beams become narrower, which is essential to attain a high \gls{snr}. In addition, the ultra-narrow beams present in \gls{fso} require high pointing precision and fast repointing to maintain adequate link quality. 

A different approach made possible by recent advances in antenna technology is the use of  phased antenna arrays, even in small satellites. In these, the antenna elements are separated by a small distance $d_e$, which is proportional to the wavelength $\lambda$, and can be used to produce highly directed beams. This enables efficient interference management due to beamforming, which exploits the spatial domain via Spatial Division Multiple Access (SDMA) or Rate-Splitting Multiple Access (RSMA), and thus, allow for an efficient use of the bandwidth. Furthermore, these beams can be steered electronically by manipulating the input signals to the antenna elements through variable phase shifters. 

Let us consider a satellite $u$ equipped with a $K\times K$ antenna array that attempts to steer the beam towards satellite $v$ at a given time $t$. To do so, it first needs to calculate the $K$-dimensional steering vectors for the azimuth angle $\phi_t^{(u,v)}$ as
\begin{IEEEeqnarray}{c}
  \mathbf{a}^{(u,v)}_{t,\text{az}}=\left[1, e^{\frac{-j2\pi d_e}{\lambda}\sin\left(\phi_t^{(u,v)}\right)},\dotsc, e^{\frac{-j2\pi d_e(K-1)}{\lambda}\sin\left(\phi_t^{(u,v)}\right)}\right]^\intercal
\end{IEEEeqnarray}
and for the polar angle $\Theta_t^{(u,v)}$ as
\begin{IEEEeqnarray}{c}
 \mathbf{a}^{(u,v)}_{t,\text{pol}}\!=\!\left[1, e^{\frac{-j 2\pi d_e}{\lambda}\cos\left(\Theta_t^{(u,v)}\right)}\!,\dotsc , e^{\frac{-j2\pi d_e(K-1)}{\lambda}\cos\left(\Theta_t^{(u,v)}\right)}\right]^\intercal\!\!.
\end{IEEEeqnarray}
Then, it calculates the overall steering vector $
    \mathbf{a}_t^{(u,v)} = \mathbf{a}^{(u,v)}_{t,\text{pol}} \otimes \mathbf{a}^{(u,v)}_{t,\text{az}}\,.
$
This approach is often called digital beam steering and it is attractive to combat the fast orbital velocities of the \gls{ngso} satellites due to its precision and switching velocity \cite{Su2019}. Nevertheless, it has the main downside that the implementation of the variable phase shifters adds a considerable amount of complexity to the hardware, which might be restrictive for nano-satellites and CubeSats.

 Butler matrix beamforming networks offer a simpler mechanism to point the beams and, hence, have gained relevance in terrestrial communications~\cite{Chang2010,Yu2018}. These are cost-efficient and low-complexity beam switching networks that produce a series of beams in pre-defined directions~\cite{Zooghby2005, Wang2018}. In contrast to digital beam steering, the beams in a Butler matrix are switched by simply feeding one or more of the fixed phase shifters (input ports), which offers an interesting trade-off between performance, cost, and complexity of operation and implementation that is especially attractive for CubeSats, which oftentimes rely on small and simple dipole antennas with low directivity. 
 
\begin{figure}[t]
    \centering
    \includegraphics{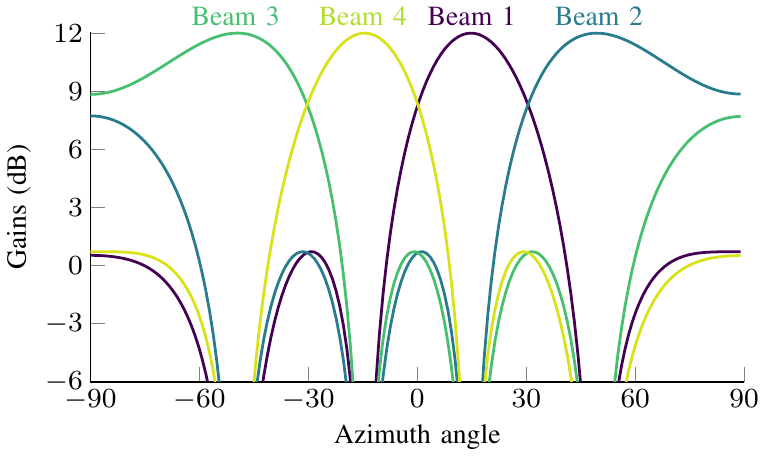}
    \caption{Gains for the beams in a $4\times4$ antenna array with Butler matrix.}
    \label{fig:gains_butler}
\end{figure}

 In particular, the steering vector in the polar angle of a Butler matrix is fixed to a specific direction $\theta$ 
 \begin{equation}
    \mathbf{b}_{\text{pol}} = \frac{1}{\sqrt{K}}\left[1, e^{\frac{-j2\pi d_e}{\lambda}\cos\left(\theta\right)},\dotsc , e^{\frac{-j2\pi d_e(K-1)}{\lambda}\cos\left(\theta\right)}\right]^\intercal,
    \label{eq:butler_altitude}
\end{equation}
whereas the steering vector of the $k$-th beam in the azimuth angle is set to 
\begin{equation}
    \mathbf{b}_{k,\text{az}} = \frac{1}{\sqrt{K}}\left[1, e^{\frac{-j\pi (2k-1)}{K} }, \dotsc , e^{-j\frac{\pi (2k-1)(K-1)}{K}}\right]^\intercal \,.
    \label{eq:butler_azimuth}
\end{equation}
The overall steering vector for beam $k$ is $\mathbf{b}_{k}=\mathbf{b}_{\text{pol}}\otimes\mathbf{b}_{k,\text{az}}$.
Fig.~\ref{fig:gains_butler} illustrates the gain of $K=4$ beams in a Butler matrix with $4\times 4$ antenna elements.

Finally, while achieving direct \gls{iot} communication with \gls{ngso} is feasible with \gls{lpwan} technologies, the use of gateways is often beneficial. These gateways may incorporate traditional dish antennas or phased antenna arrays that gather the transmissions from \gls{iot} devices with non-directive antennas and, then, transmit to the satellites with highly directive antennas. However, another option made possible by the predictable movement of the satellites is to deploy \glspl{irs}. These low-complexity elements that modify the characteristics of the incident signals and, hence, can help direct the signals towards the satellites~\cite{Matthiesen2021}.

\subsection{Frequent link establishment and adaptation}
\label{sec:link_establishment}

Due to the movement of the satellites, the physical links must be frequently re-established and adapted. This includes selecting the pairs of satellites to establish the \glspl{isl}, beam pointing/steering or switching for the Butler matrix case, and rate adaptation. 
Since the movement of the constellation is fully predictable, these problems can be solved in advance with a specific optimisation objective in mind. These objectives depend on the target service(s) and can be, as listed in Section~\ref{sec:intro}, to maximize the \emph{transport capacity}~\cite{Liu2017, Jiang2020} of the constellation or to minimize the \gls{e2e} latency for a set of specific paths. 

Some constellations designs are fully symmetric, with each and every one the orbital planes containing the same number of satellites and these being deployed at the exact same altitude. In these cases, the orbital period $T_o$ of all the satellites is exactly the same and, hence, these will all be periodically at the exact same position. In these cases, the optimal configuration of the links can be obtained for several instants within the period $T_o$ and applied periodically.

However, asymmetries in the constellation are usually present, either 1) to enhance the sustainability of the constellation by considering orbital separation as in OneWeb~\cite{Lewis2019}, 2) to fulfill certain coverage and service availability targets by incorporating several orbital shells as in Starlink, or 3) to provide service during the initial phases of deployment of the constellation. In these cases, fixed solutions cannot be used and the links must be established on the fly.

An essential aspect for link establishment and maintenance is to implement an adequate beam steering technology as discussed in the previous section. Furthermore, the \gls{mcs} and transmit power may be adapted to maximize throughput and reliability while minimizing potential interference. Naturally, the characteristics of the antennas and beams must be considered during link establishment~\cite{Leyva-Mayorga2021GC}.

An option to re-establish the links is to treat the link establishment as a one-to-one matching problem in a dynamic weighted graph $\mathcal{G}_t=\left(\mathcal{V},\mathcal{E}_t\right)$, where the satellite antennas, transceivers, or even beams (for the case of beam selection) form a multi-partite vertex set $\mathcal{V}$ and the weighted edge set at time $t$, denoted as $\mathcal{E}_t$ are the feasible \glspl{isl} with non-zero rates. Then, the matching at a time $\mathcal{M}_t$ is the set of pairs of antennas/transceivers/beams and the rates for communication. In this case, the matching $\mathcal{M}_t$ can be calculated periodically, once every $\Delta t$ seconds, in a centralised entity with full knowledge of the constellation parameters and dynamics. Then, the solution for the matching for the satellite positions at time $t$ must be propagated through the constellation before this time. With the full predictability of the movement of the constellation, the solution can be calculated sufficiently in advance and, hence, the latency of communicating it to the satellites is irrelevant. Hence, this approach can lead to near-optimal or optimal solutions at the expense of injecting periodic traffic into the network to communicate the solution to all the satellites. An important aspect of the inter-plane \gls{isl} link establishment is that the graph $\mathcal{G}$ that represents a single orbital shell is multi-partite, with each subset representing an orbital plane and, hence, traditional algorithms such as the Hungarian algorithm cannot be used and other solutions are needed.

On the other hand, localised decisions may be implemented, for example using distributed algorithms for the matching. An example of these is the Deferred Acceptance algorithm~\cite{Gu2015}, where the individual agents maintain and inform their preferences to the neighbourhood and the matching is solved in parallel, after a few iterations. While more research is needed to determine the performance of distributed vs. centralised matching solutions, distributed algorithms are required 1) to establish the links during the deployment phase before the constellation is fully operative, and 2) in case the connection with the centralised entity is lost.

To solve the inter-plane \gls{isl} establishment problem, we have explored the use of greedy matching algorithms with 2) ideal beam pointing (i.e., at each time $t$) and resource allocation and 2) with periodic repointing via digital beamforming and beam switching via Butler matrix beamforming networks with period $\Delta t$~\cite{Leyva-Mayorga2021},~\cite{Leyva-Mayorga2021GC}. Algorithm~\ref{alg:matching} illustrates the steps of a general greedy matching algorithm for link establishment. The latter can be extended to include orthogonal resource allocation (e.g., frequency sub-bands) to minimize interference~\cite{Leyva-Mayorga2020}. 

\begin{algorithm} [t]
	\centering
	\caption{Greedy satellite matching with multiple beams.}
	\renewcommand{\arraystretch}{1.3}
	\begin{algorithmic}[1] 
	\renewcommand{\algorithmicrequire}{\textbf{Input:}}
		\renewcommand{\algorithmicensure}{\textbf{Output:}}
		\REQUIRE Set of feasible weighted edges $\mathcal{E}_t$ and $\mathcal{E}_{t+\Delta t}$ and the initial state of the matching $\mathcal{M}_t$
		\REQUIRE Antenna configuration
		\STATE Initialise indicator variables
		\WHILE {More edges can be matched}
		\STATE Select the edge with maximum weight
		\IF {the selected vertices are not in $\mathcal{M}_t$}
		\STATE Add the vertices to the matching $\mathcal{M}_t$
		\STATE Update the indicator variables
		\STATE Remove all adjacent edges to the selected vertices from $\mathcal{E}_t$ and $\mathcal{E}_{t+\Delta t}$
		\STATE Update the interference and weights to all edges in $\mathcal{M}_t$, $\mathcal{E}_t$ and $\mathcal{E}_{t+\Delta t}$
		\ENDIF		
		\ENDWHILE
		\end{algorithmic}  
	\label{alg:matching}
\end{algorithm}

\begin{figure}[t]
    \centering
    \subfloat[]{\includegraphics{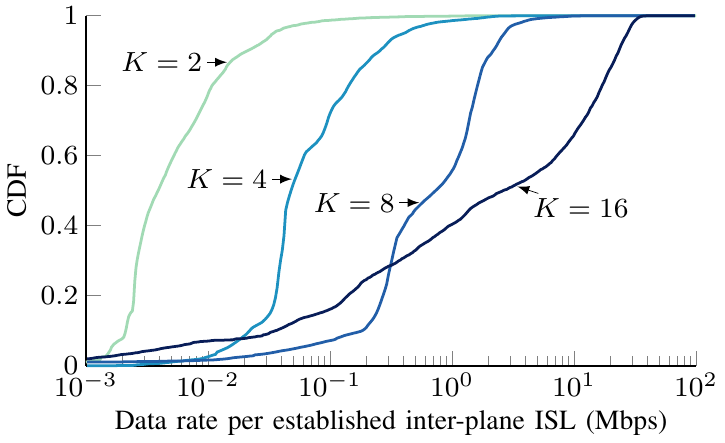}\label{fig:rates_isl_butler}}\hfil
    \subfloat[]{\includegraphics{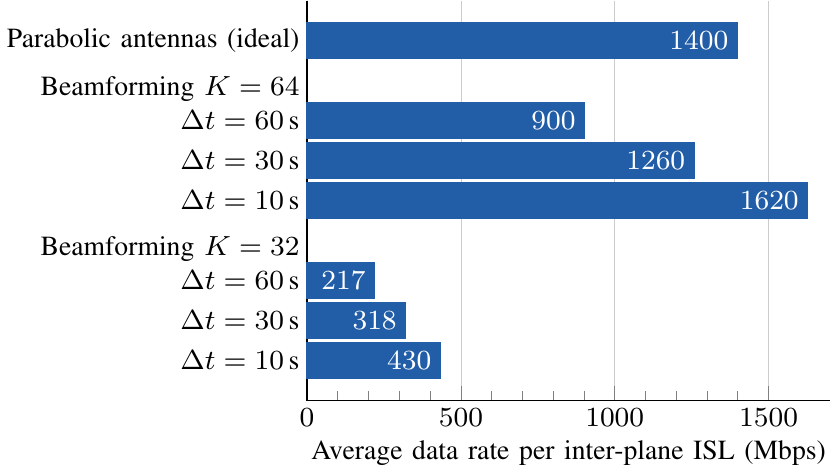}\label{fig:rates_isl_beamforming}}
    \caption{(a) \gls{cdf} of the rates per inter-plane \gls{isl} with Butler matrix arrays and (b) average rates per inter-plane \gls{isl} with parabolic antennas with ideal pointing and for digital beam forming for different link re-establishment periods $\Delta t$.}
    \label{fig:rates_isl}
\end{figure}

Note that Algorithm~\ref{alg:matching} attempts to maximize the sum of weights in the matching. Throughout our previous work, we have defined the weights to be the achievable rate for communication at the \glspl{isl} in $\mathcal{E}_t$. Following this approach, Fig.~\ref{fig:rates_isl_butler} illustrates the increase of the rates per \gls{isl} as a function of the number of elements $K$ in a Butler matrix beamforming network for the Kepler constellation. As it can be seen in Fig.~\ref{fig:rates_isl_butler} increasing the number of elements $K$ greatly improves the data rates, however, this also increases the number of beams that must be considered by the matching algorithm and, hence, increases the running time of the algorithm. 

Furthermore, Fig~\ref{fig:rates_isl_beamforming} shows the effect of the re-establishment period $\Delta t$ on the average data rate per inter-plane \gls{isl} with digital beamforming; the data rate achieved with ideal pointing (i.e., with $\Delta t = 0$ and parabolic antennas is included as a reference. It can be seen that increasing the frequency of link re-establishment and adaptation increases the data rates and phased antenna arrays of $K=64$ can be used to achieve similar rates as with greatly directional parabolic antennas, even with ideal pointing. However, the re-establishment period cannot be reduced arbitrarily as this can cause problems, for example, for routing algorithms, due to the frequent changes of the network topology.
 
Throughout our analyses, we have observed that Butler matrix networks with relatively low dimensions $K$ are an attractive option for the inter-plane link establishment in resource-constrained satellites (i.e., CubeSats and small-sats). However, if large antenna arrays and variable phase shifters can be implemented on the satellites, beamforming offer gains in the data rates that are greater than $200$\% and, hence, these should be preferred.

It is important to mention that rate maximisation does not directly increases the \emph{transport capacity} f the network, which is a difficult measure to define. Usually, specific source-destination pairs are defined and the transport capacity is the maximum amount of data (i.e., flow) that can be transmitted between them~\cite{Ahlswede2000}. In these cases, calculating the transport capacity usually involves assigning flow to all possible paths from the source to the destination, as in the Edmons-Karp algorithm, which has been used to calculate the capacity of constellations with multiple orbital shells~\cite{Jiang2020}. However, this is complicated in dynamic and large networks, so upper bounds based on selecting cuts from the network graph have been used~\cite{Liu2017}.
Yet another hindrance of using the Edmonds-Karp algorithm is that is assumes that ideal mechanisms to redistribute the traffic flows are in place. Instead, in a network with multiple source-destination pairs, the capacity of some links is likely to be shared among them and the number of alternate paths may be limited depending due to the implemented routing, load balancing, and congestion control mechanisms. Furthermore, the distribution of the traffic among the different paths will usually be imbalanced. Therefore, defining the transport capacity of a satellite constellation is complicated. 

A simple scenario where it is possible to calculate the maximum [G2G] traffic that can be generated from each \glspl{gs} is where these have equal traffic characteristics and where unipath source routing is used~\cite{rabjerg2021routing}. In this scenario, we can define $\mathcal{P}_t$ as the set of possible paths at time $t$. A path $\mathbf{p}\in \mathcal{P}_t$ is an ordered set of edges, denoted as $\mathcal{E}(\mathbf{p})=\left(e_1,e_2,e_3,\dotsc\right)$. Here, the load $\lambda$ of each of the $N_\text{GS}$ \glspl{gs} is distributed evenly to the rest of the $N_\text{GS}-1$ \glspl{gs}, using the paths in $\mathcal{P}_t$. Hence, the load assigned to each path $\mathbf{p}\in \mathcal{P}_t$ is 
\begin{equation}
    \lambda_\mathbf{p} = \frac{2\lambda}{ N_{\text{GS}} - 1}.
\end{equation}
The max-flow min-cut theorem states that the maximum flow that can be transmitted through a path is determined by the link (i.e., edge) with minimum capacity (i.e., throughput)~\cite{Ahlswede2000}. Hence, at a time $t$ we have
\begin{equation}
    \sum_{\mathbf{p}\in \mathcal{P}_t}~\sum_{uv\in \mathcal{E}_t(\mathbf{p})} \lambda_\mathbf{p}= N_{\mathbf{p}}(uv)\lambda_{\mathbf{p}}\leq R_t(u,v), \quad \forall u,v\in \mathcal{V},
\end{equation}
where $N_\mathbf{p}(uv)$ is the number of paths in $\mathcal{P}_t$ that contain the edge $uv$. Naturally, $N_{\mathbf{p}}(uv)$ depends on the routing metric.
Building on this, it is possible to calculate the maximum load per \gls{gs} at time $t$ as
\begin{equation}
    \lambda_t^*=\min_{uv\in\mathcal{E}_t}\frac{R_t(u,v)\left(N_{\text{GS}} - 1\right)}{N_{\mathbf{p}}(uv)}. \label{eq:lambdamax}
\end{equation}

\subsection{Routing, load balancing and congestion control}
A general goal to achieve in the design of higher layer algorithms is to account for both the traffic characteristics (load, queues, and QoS/QoE requirements) and the instantaneous state of the links/paths. However, the time-variations of the traffic and the channel are different in terrestrial and satellite networks and, e.g.,
the conventional TCP/IP stack is ineffective against the long delays, packet losses and intermittent connectivity that characterizes \gls{ngso} communications. Therefore, specific networking solutions are required.  %In the case of NGSO constellation, the terrestrial solutions need adaptation to take into consideration the time scales and the predictable movements. %This requires to frequently update the data rates and the transmission queues at each of the links, which allows to select the best route to the destination.
%\TDAA{Expand on high-layer functionalities}

A routing algorithm is a collaborative process for deciding, in every intermediate node, the directions to reach the destination as soon as possible.\footnote{In NGSO constellations and other satellite systems, a second option for delay-tolerant applications is the store-carry-forward strategy where nodes can temporarily store and carry in-transit data until a suitable link becomes available, e.g., until the next pass with a ground station.}
This routing problem presents the following unique characteristics in NGSO constellations:
\begin{itemize}
\item The topology is highly dynamic, with frequent handovers in the links between ground and NGSO, and between NGSOs in different orbital planes (inter-plane ISL).  
\item The load from the ground terminals (ground stations and users) is imbalanced, with 1) some satellites serving, e.g., deserted/ocean areas while other nodes pass above densely populated regions and 2) some source-destination pairs experiencing more intense data flows than others.
\item The need to have a reliable and resilient routing solution, which implies that the satellite segment must possess a sufficient degree of autonomy to cope with, e.g., queueing delays or local link or satellite failures and find alternative routes at each time instant. However, this must be achieved while exchanging minimal feedback and routing information to limit the signaling overhead. 
\end{itemize}

A good overview of routing protocols for satellite can be found in ~\cite{ruiz2018routing}. Most previous works have oversimplified the ground/space segments geometry and the ISL connectivity to focus on other challenges like the QoS. One prominent exception is \cite{Handley2018}, although the study is for a specific commercial constellations. \cite{rabjerg2021routing} takes a more general approach and focuses on two distinctive elements to the routing problem in a NGSO constellation. First, the propagation time has a great impact on the overall latency, contrary to terrestrial mesh networks. Second, the location of the ground stations greatly impacts the traffic load injected to the constellation and the geographic locations where the traffic is injected. As in  Section~\ref{sec:link_establishment}, the space and ground infrastructure at a given time $t$ is modeled as a dynamic weighted undirected graph $\mathcal{G}_t=\left(\mathcal{V}, \mathcal{E}_t\right)$. However, by adding the \glspl{gs}, the vertex set is now defined as $\mathcal{V}=\mathcal{U}\bigcup_{a\in\mathcal{P}} \mathcal{V}_a$, where $\mathcal{U}$ is the set of ground stations and $\mathcal{V}_a$ is the set of satellites deployed in orbital plane $a$, and $\mathcal{P}=\{1,2,\dotsc,P\}$ is the set of orbital planes. The edge set $\mathcal{E}_t$ represents the wireless links available for communication. For instance, the satellites might deploy four ISLs at all times: two intra-plane ISLs and two inter-plane ISLs. In this case, the intra-plane ISLs within an orbital plane $a$ constitute the fixed set of edges $\mathcal{E}^{(a)}=\left\{uv:u,v\in\mathcal{V}_a\right\}\subset \mathcal{E}_t$. On the other hand, the inter-plane \glspl{isl} between orbital plane $a$ and orbital plane $b$ constitute the set of edges $\mathcal{E}^\text{inter}_t=\left\{uv:u\in\mathcal{V}_a, v\in\mathcal{V}_b, a\neq b\right\}\subset \mathcal{E}_t$; as mentioned in Section~\ref{sec:link_establishment}, these must be frequently re-established due to the movement of the satellites. Furthermore, the ground stations maintain one \gls{gsl} with their closest satellite at all times. These \glspl{gsl} constitute the set of edges 
\[\mathcal{E}^G_t = \left\{uv:u\in\mathcal{U}, v\in\mathcal{V}_a, a\in\mathcal{P}\right\}.\]
Finally, we define the edge set as
\[\mathcal{E}_t=\mathcal{E}^G_t\cup\mathcal{E}^\text{inter}_t\bigcup_{a\in\mathcal{P}}\mathcal{E}^{a}_t.\]
 
The route of a single packet transmitted at time $t$ is then a weighted path $\mathbf{p}$ in $\mathcal{G}_t=(\mathcal{V},\mathcal{E}_t)$ with edge set $\mathcal{E}(\mathbf{p})$. The weights $w(e)$ for all $e\in \mathcal{E}_t$ are defined by the routing metric to account to, e.g., the path loss and/or the communication latency. Specifically, capturing the non-linearity of the path loss in the ISL will favour paths with high-data rates and consequently reduce the waiting times in the buffers. Rather than complex feedback mechanisms to collect up-to-date network status information, this simpler approach has proven to provide a good trade-off between complexity and performance. 

The degree of integration of the \gls{ngso} constellation with the terrestrial infrastructure has also an effect on the traffic load. Not in vain, a prominent application of 5G satellite communications is to offload the terrestrial networks in congested urban areas, either with direct satellite access or through a gateway~\cite{soret2020backhauling}. In both cases, it will further exacerbate the load imbalance. A subsidiary case is the use of the constellation as a backhaul that
transparently carries the payload between the two communication extremes. This is typically used to connect isolated base stations to the core network. 

Regarding resilience, the classical approach to space routing is to centrally compute all the paths in a location register, and then broadcast the information to all the satellites. Satellites forward the packets according to the on-board routing tables, which are configured based on the central computations. In the case of \gls{ngso}, the central location register can be a terrestrial station or a \gls{geo} satellite. In any case, this approach scales poorly due to the highly dynamic topology, with frequent handover events between nodes and terminals causing significant signaling overhead. Moreover, the current status of the satellites (load, buffers, batteries) should be included in the decision, but this requires an enormous amount of feedback from each node in the graph to the central location register. The alternative is to move towards more distributed solutions. From semi-distributed to fully autonomous algorithms, the idea is that each satellite decides the next hop for each received packet, taking into consideration all the available information, including the prior knowledge (past) and the predicted paths (future). 

As in terrestrial networks, the space network might be shared by several services with heterogeneous requirements. For example, some broadband users require high rates, as provided by the \gls{geo} segment, whereas IoT devices are sensitive to delays or freshness of the information~\cite{soret2020backhauling}, better provided by the \gls{ngso} segment, or some services demand extra satellite computation. In general, there are multiple paths for most source-destination pairs, and this diversity should be exploited to meet the heterogeneity of requirements.

\begin{figure}[t]
\centering
\subfloat[]{\includegraphics{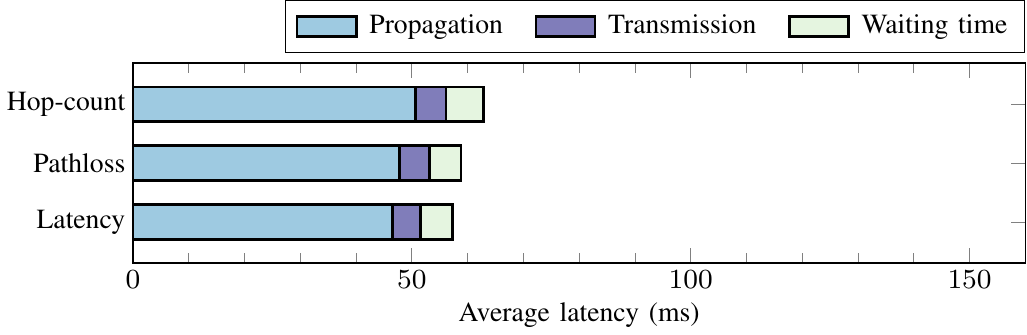}}\\
\subfloat[]{\includegraphics{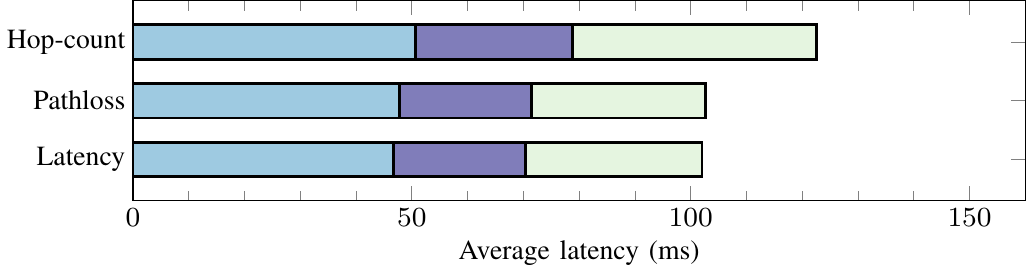}}\\
\subfloat[]{\includegraphics{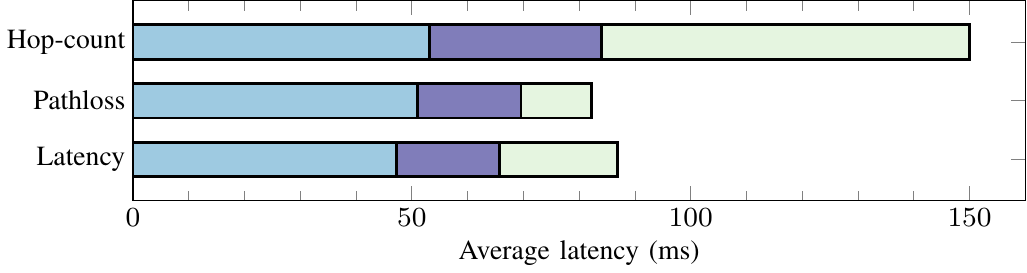}}
\caption{\rev{Average routing latency per packet due to propagation, transmission, and waiting times for three topology-aware metrics with: (a) Kepler constellation; (b) Kepler constellation with $P_t=1$ W; and (c) constellation with $P=5$, $N_\text{op}=40$, and $P_t=1$ W.}}
\label{fig:routing_latency}
\end{figure}

The example in Figure~\ref{fig:routing_latency} illustrates the performance of different routing metrics, taking the latency as the \gls{kpi} of interest. \rev{Three different configurations are considered: (a) the Kepler constellation with the communication parameters listed in Table~\ref{tab:comm_params}; (b) the Kepler constellation with transmission power $P_t=1$ W; and (c) a Walker-star constellation with $P=5$ orbital planes at height $h=600$ km and $N_\text{op}=40$ satellites per orbital plane and $P_t=1$ W}. The time-varying data rate follows the channel variations. The ground segment consists of $N_{\text{GS}} = 23$ \glspl{gs} placed accordingly to the KSAT ground station service\footnote{\href{https://www.ksat.no/services/ground-station-services/}{https://www.ksat.no/services/ground-station-services/}. The details of the simulations can be found in~\cite{rabjerg2021routing}.} 

\rev{The compared metrics are: (1) a classical hop-count approach that merely minimizes the number of hops to reach the destination; (2) a path loss metric that considers the non-linearity of the \gls{isl}; (3) a latency metric that takes into account the propagation and transmission times but skips the need for a feedback channel by having a statistical model of the queueing times. As expected, the latency metric effectively selects the routes with the shortest propagation and transmission times in all three cases. However, the waiting times are shorter with the pathloss metric. This is because the pathloss metric emphasises the selection of high data rate links over short routes, which support greatest traffic load. As a consequence, the pathloss metric leads to the  lowest overall latency with configuration (c) and to a closely similar latency to the latency metric in the other two cases. The reason for this is that the configuration with $P=5$ and $N_\text{op}=40$ has a greater density of satellites along the orbital planes, which leads to a much greater data rate at the intra-plane when compared to the inter-plane \glspl{isl}. These links are prioritized by the pathloss metric. Furthermore, it can be observed that, while the propagation delay changes slightly, the choice of communication and constellation parameters greatly affects the transmission and waiting times. Finally, Figure~\ref{fig:routing_latency} illustrates the need for an advanced routing metric: even when the number of satellites with configuration (c) is greater than that with the other two configurations, the latency achieved by the hop-count metric is greater for this case.}

A complementary function to routing is congestion control, which aims at ensuring high bandwidth utilisation while avoiding network congestion. This is done at the transport layer by regulating the rate at which traffic sources inject packets into the network. However, the standard TCP assumes that the bottleneck link will stay the same over time and that changes in its capacity are erratic. This is not true in the satellite network case, in which the capacity of a link is predictable and therefore a location-aware congestion control mechanism can improve the throughput and latency. In this direction, several works have proposed variations of TCP for space networks. Although the topic is definitely not new~\cite{Durst1997}, the initial works were targeting space networks very different from \gls{ngso} constellations, where delay- and disruption-tolerant satellite applications and large distances Earth-GSO were the norm. For example, the Space Communications  
Protocol Specification-Transport Protocol (SCPS-TP), mainly developed by NASA and the US Department of Defence, has a selective negative acknowledgement to accommodate asymmetric channels and explicit congestion notification~\cite{Durst1997}. Another option that does not modify the underlying protocol is the \gls{dtn} architecture, which provides long-term information storage on intermediate nodes to cope with disrupted or intermittent links~\cite{Caini2011}. 
A more recent alternative is the use of QUIC (Quick UDP Internet Connections), the general purpose transport protocol defined by Google~\cite{hamilton2013} to combine the advantages of connected-oriented TCP and low-latency UDP. \gls{ngso} networks can benefit from QUIC~\cite{siyu2018} when there is a high \gls{rtt} and a poor bandwidth. Moreover, QUIC introduces a connection ID instead of IP addresses as identification which inherently avoids re-connections with frequently changeable topological space networks. 

%%%%%%%%%%%%%%%%%%%%%%%%%%%%%%%%%%%%%

\section{Conclusions}
In this chapter, we described relevant aspects of \gls{ngso} constellation design to achieve global connectivity. That is, to provide global service availability to ground terminals but also to ensure inter-satellite connectivity can be achieved along the constellation. We emphasized that the constellation geometry, the altitude of deployment, and the density of satellites have a major impact on these and other relevant \glspl{kpi} and compared the performance of three commercial designs: Kepler, OneWeb, and the Starlink orbital shell at $550$\,km. \rev{We observed that, while the Starlink orbital shell has a greater number of satellites than the other two constellations, it still requires an additional orbital shell with nearly-polar orbital planes to provide connectivity near polar regions. On the other hand, around $45$ satellites from the OneWeb constellation are simultaneously within communication range in the near-polar regions, which may lead to waste of communication resources. Finally, the Kepler constellation may suffer from coverage holes near the Equator where, on average, less than one satellite is within communication range from the Earth's surface. To provide ubiquitous global coverage, a constellation similar to Kepler but with slightly larger number of orbital planes and satellites would be sufficient. Still, NGSO constellations that aim to provide broadband services would benefit from further increasing the density of deployment, which would lead to greater data rates both in the inter- and intra-plane \gls{rf} \glspl{isl}.} 

\rev{Besides the impact of the main parameters for constellation design,} we elaborated on the major challenges and technologies to achieve global connectivity at the physical layer, for link establishment, and routing. These arise from the distinctive characteristics of \gls{ngso} constellations, which are greatly dynamic, yet fully predictable large-scale infrastructures.

\section*{Acknowledgements}
This work is supported in part by the German Federal Ministry of Education and Research (BMBF) within the project Open6GHub under grant number 16KISK016 and by the German Research Foundation (DFG) under grant EXC 2077 (University Allowance).

\bibliographystyle{vancouver}

\end{document}